\documentclass{article}

 \newcommand{\ben}{\begin{enumerate}}
\newcommand{\een}{\end{enumerate}}
\newcommand{\beq}{\begin{equation}}
\newcommand{\eeq}{\end{equation}}
\newcommand{\bse}{\begin{subequation}}
\newcommand{\ese}{\end{subequation}}
\newcommand{\bea}{\begin{eqnarray}}
\newcommand{\eea}{\end{eqnarray}}
\newcommand{\bc}{\begin{center}}
\newcommand{\ec}{\end{center}}

\def\DR{\rm I\kern-1.45pt\rm R}
\def\DC{\kern2pt {\hbox{\sqi I}}\kern-4.2pt\rm C}
\def\DH{\rm I\kern-1.5pt\rm H\kern-1.5pt\rm I}
\begin{document}

\begin{center}
{\Large\bf The Stark effect in the charge-dyon system}\\[3mm]
{\large Levon Mardoyan${}^{1,3}$, Armen Nersessian${}^{1,2}$, 
Mara Petrosyan${}^{1,3}$}
\end{center}
{\it
${}^{1}$ International Center for Advanced Studies, Yerevan State
University, Yerevan\\
${}^{2}$ Yerevan Physics Institute, Yerevan \\ 
${}^{3}$University of Nagorny Karabakh, Stepanakert}

\begin{abstract}
 The linear Stark effect in the MIC-Kepler problem describing the
interaction of charged particle with Dirac's dyon is considered.
It is shown that constant homogeneous electric field completely
removes the degeneracy  of the energy levels on azimuth quantum
number.
\end{abstract}

\vspace{0.5cm}

In February 2003 Professor Valery Ter-Antonyan: an intellectual,
brilliant pedagogue and the expert in  quantum
mechanics,  passed away. During the last years of his life he
took a great interest in the study of the Coulomb problem in the
presence of topologically nontrivial objects. The two of us had
the honour of collaborating with him in this sphere. This modest
work is a tribute to his memory.

\section{Introduction}

The integrable system MIC-Kepler  was constructed by Zwanziger
\cite{Z} and rediscovered by McIntosh and Cisneros \cite{mic}.
This system is described by the Hamiltonian
\begin{equation}
{\cal H}_0=\frac{\hbar^2}{2\mu}(i{\bf{\nabla}}+ {s}{\bf A})^2
+\frac{\hbar^2{s}^2}{2\mu r^2}-\frac{\gamma}{r},\quad{\rm
where}\quad {\rm rot}{\bf A}=\frac{{\bf r}}{r^3}. \label{1}
\end{equation}
Its distinctive peculiarity is the  hidden symmetry given
by the following constants of motion (\ref{1})
\begin{equation}
{\bf I}=\frac{\hbar}{2\mu}\left[(i{\bf\nabla}+ {s}{\bf
A})\times{\bf J}-{\bf J}\times (i{\bf\nabla}+ {s}{\bf A})\right]
+\gamma\frac{{\bf r}}{r},\quad {\bf J}=-\hbar (i{\bf\nabla}+
{s}{\bf A})\times{\bf r}
 +\frac{\hbar{s}{\bf r}}{r}.
\label{2}
\end{equation}
These constants of motion, together with the Hamiltonian, form the
quadratic symmetry algebra of the Coulomb problem.
The operator  ${\bf J}$ defines  the angular momentum of the
system, while the operator ${\bf I}$ is  the analog of the Runge-Lenz
vector.
For the  fixed negative values of energy  the constants of motion
form  the $so(4)$ algebra, whereas for positive values of energy - 
the  $so(3.1)$ one. Due to the hidden symmetry  the MIC-Kepler problem 
could be factorized in few coordinate systems, e. g. in 
the spherical and  parabolic ones.
Hence, the MIC-Kepler system is a natural generalization of the Coulomb
problem in the presence of Dirac monopole. The
monopole number $s$ satisfies the Dirac's  charge
quantization rule,  $s=0,\pm1/2,\pm 1,\ldots$.

The MIC-Kepler system could be constructed by the reduction of the
four-dimensional isotropic oscillator by the use of the so-called
Kustaanheimo-Stiefel transformation both on classical and quantum
mechanical levels \cite{nt}. In the similar way, reducing the two-
and eight- dimensional isotropic oscillator, one can obtain the
two-  \cite{ntt}  and five-dimensional  \cite{iwa1} analogs of
MIC-Kepler system. An infinitely thin solenoid providing the
system by the spin $1/2$, plays the role of monopole in
two-dimensional case, whereas in the five-dimensional case this
role is performed by the $SU(2)$ Yang monopole \cite{yang},
endowing the system by the isospin. All the above-mentioned
systems have Coulomb symmetries and are solved in spherical and
parabolic coordinates both in discrete and continuous parts of
energy spectra \cite{mardoyan}. There are generalizations of
MIC-Kepler systems on three-dimensional sphere \cite{kurochkin}
and hyperboloid  \cite{np} tas well.

For integer values $s$ the MIC-Kepler system describes the
relative motion of the two Dirac dyons (charged magnetic
monopoles). For this we should place
\begin{equation}
s=\frac{eG-Qg}{\hbar c},\quad \gamma=eQ+gG,
\label{AS}
\end{equation}
where $(e, g)$ and $(Q, G)$ are electric and magnetic charges of
the first and the second dyons respectively.
 Parameter $\mu$ plays
the role of the reduced mass, while vector ${\bf r}$ determines
the position of the second dyon with respect to the first one
\cite{Z}.
 For half-integer $s$ the presence of the 
magnetic field of infinitely thin solenoid,
 endowing the system with the spin $1/2$, is also supposed (compare
 with  \cite{ntt}).
 It should be noted that the
MIC-Kepler system describes also the relative motion of two Dirac 
dyons as well as that of the twowll-separated BPS monopoles/dyons
\footnote{The system describing the relative motion of the two well-separated
BPS dyons \cite{gm} is specified by the Hamiltonian 
\beq 
{\cal H}_{BPS}=\frac{1}{4\mu}
\left(1-\frac{2\mu}{r}\right)^{-3/2}(i{\bf{\nabla}}+ {s}{\bf
A})\left(1-\frac{2\mu}{r}\right)^{1/2}(i{\bf{\nabla}}+ {s}{\bf
A})+\frac{eg}{r},\quad \eeq
where $e$ denote the relative electric
charge of the BPS monopoles. As is seen the Schr\"{o}dinger
equation of this Hamiltonian could be immediately
transformed to the
Schr\"{o}dinger equation of the MIC-Kepler problem. While the 
 Coulomb symmetry in this system is well-known one, its equivalence to 
the MIC-Kepler problem might have been noticed in \cite{bn}}.

The wavefunction of the ground state of the MIC-Kepler problem is
of the  form:
\beq \psi_{m,\pm |s| }={\rm const}\; r^{|{s}|}
e^{-r/a(|{s}|+1)}(\cos\frac{\theta}{2})^{|{s}|\pm m}
(\pm\sin\frac{\theta}{2})^{|{s}|\mp m}e^{im\varphi},\quad
m=-|s|,-|s|+1,\ldots, |s|-1,|s|. \label{gs} \eeq 
As is seen, it is degenerated  with respect to the quantum number $m$ 
and is not spherically symmetric: the 
system has a non-zero dipole momentum and 
in the presence of the external
electric field the linear Stark effect is possible. 
 Hence,  it seems to be interesting to study the behavior of the 
MIC-Kepler bound system behaviour in the constant uniform
electric field with the purpose to  investigate the Stark effect
and to calculate  the dipole momentum. The aim of the
present paper is the study of these issues.

For convenience, we will consider a special case of the charge
moving  in the field of the Dirac dyon, i.e. the system most  symilar 
 to the hydrogen atom. 
In other words, we assume 
\beq G=0,
\label{chg} \eeq
 so that the perturbative correction to the
MIC-Kepler Hamiltonian (\ref{1}), which is  responsible for 
the interaction of
the charge-dyon system with the external constant uniform electric field
has the same form as in the case of the hydrogen atom,
\beq
{\cal H}_S= e{{\mbox{\boldmath $\varepsilon$}}}{\bf r}.
\eeq

The  paper is organazed as follows.

In Section 2 we bring wave functions of the MIC-Kepler bound
system in spherical and parabolic coordinates.

In Section 3 the behaviour of the charge-dyon bound system in the
constant uniform  electric field is investigated. The
assumption of the presence of the linear Stark effect in the
system is confirmed and its dipole momentum is calculated. It is
shown that the external electric field completely removes the
degeneracy  on azimuth quantum number.

In Conclusion we summarize the obtained results 
discuss their   possible
generalizations.

\setcounter{equation}{0}
\section{Wavefunctions and spectrum}

Let us consider a spectral problem describing the MIC-Kepler
problem with the energy ${E}^{(0)}$, with the angular momentum $j$ 
and with the   $x_3$-component of the angular momentum $m$: 
\beq 
{\cal H}_{0}\psi={E}^{(0)}\psi,\quad {\bf J}^2\psi=\hbar^2j(j+1)\psi
\quad J_3\psi=\hbar m\psi \label{sp} \eeq 
where ${\cal H}_{0},{\bf J}$
 are defined  by the expressions (\ref{1}),(\ref{2})
with the Dirac's monopole vector potential with the sigularity line
 directed along the positive semiaxis $x_{3}$
\beq
{\bf A} = \frac{1}{r(r - x_3)}\left(x_2, -x_1, 0\right)
\eeq
The solution to this system is as follows:
\begin{equation}
\psi_{njm}({\bf r} ;{s}) =
\left(\frac{2j+1}{8\pi^2}\right)^{1/2}R_{nj}({r}/{a})
d^j_{ms}(\theta)e^{im\varphi},\label{wfs}
\end{equation}
where $d^j_{ms}$ is the Wigner d-function \cite{11}, and
$R_{nj}({r}/{a})$ is defined  by the expression 
\begin{eqnarray}
R_{nj}(r)& =& \frac{2^{j+1}}{n^{j+2}(2j+1)!}
\sqrt{\frac{(n+j)!}{(n-j-1)!}}r^j e^{-r/n} F(j-n+1,
2j+2,\frac{2r}{n}).
\end{eqnarray}
Here $a=\hbar^{2}/\mu e^{e}$ is Bohr radius. The spectrum of the
system is specified by the conditions:
\begin{eqnarray}
&{E}^{(0)}_n = -\frac{\mu \gamma^2}{2\hbar^2n^2}, n=|s|+1,|s|+2,
\ldots&\label{senergy}\\
&j= |{s}|,  |{s}| +1,\ldots, n-1;\quad  m =-j, -j+1,\ldots, j-1,
j &.
\end{eqnarray}
The quantum numbers $j,m$ characterize the total momentum of the
system and its projection on the axis $x_3$. For the (half)integer $s$
$j, m$ are (half)integers.
Performing  the identity transformation $\varphi\to\varphi+2\pi$, one can see, that the
wavefunction of the system is single-valued for the integer ${s}$
and changes its signs for the half-integer ${s}$. In the latter case
 the ambiguity of the wave function can be interpreted as the
presence of the magnetic field of the infinitely thin solenoid (directed
along the axis $x_3$) providing the system by  spin $1/2$. 
In the ground state ($n=1$) we have $j=|{s}|$,
so that the wavefunction is of the  form (\ref{gs}).

Let
us consider the MIC-Kepler system on the parabolic basis.
In the parabolic coordinates 
$\xi,\eta \in [0, \infty), \, \varphi\in [0, 2\pi)$,
 defined by the formulae
\beq x_1+ix_2 = \sqrt{\xi \eta}e^{i\varphi}, \qquad x_3 =
\frac{1}{2}(\xi - \eta), \label{parabolic}\eeq
the differential elements of length and volume read
\beq
dl^2 = \frac{\xi + \eta}{4}\left(\frac{d\xi^2}{\xi} +
\frac{d\eta^2}{\eta}\right) + \xi \eta d\varphi^2, \qquad dV=
\frac{1}{4}(\xi + \eta)d\xi d\eta d\varphi, \label{pmetric} \eeq
while
the 
Laplace operator looks as follows
\beq
\Delta =\frac{4}{\xi +\eta}\left[\frac{\partial}{\partial \xi} \left(\xi
\frac{\partial}{\partial \xi}\right) + \frac{\partial}{\partial
\eta}\left(\eta \frac{\partial} {\partial \eta}\right)\right] +
\frac{1}{\xi \eta} \frac{\partial^2}{\partial \varphi^2}.
\label{laplasian}
\eeq
The
substitution
\beq \psi(\xi,\eta,\varphi) = \Phi_1(\xi)
\Phi_2(\eta)\,\frac{e^{im\varphi}}{\sqrt{2\pi}}.
\eeq
separates 
the variables in the Schr\"{o}dinger equation 
and  we
arrive to the following system 
\begin{eqnarray}
\frac{d}{d \xi}\left(\xi \frac{d\Phi_1}{d \xi}\right) +
\left[\frac{\mu E^{(0)}}{2\hbar^2}\xi - \frac{(m-s)^2}{4\xi} +
\frac{\mu}{2\hbar}\beta +
\frac{1}{2a}\right]\Phi_1 &=& 0, \\
\nonumber\\
\frac{d}{d \eta}\left(\eta \frac{d\Phi_2}{d \eta}\right) +
\left[\frac{\mu E^{(0)}}{2\hbar^2}\eta - \frac{(m+s)^2}{4\eta}-
\frac{\mu}{2\hbar}\beta + \frac{1}{2a}\right]\Phi_2 &=& 0,
\end{eqnarray}
where $\beta$ -- is the separation constant, being the eigenvalue
of the $x_3$-component the Runge-Lenz vector ${\bf I}$.

For $s=0$ these equations coincide with the equations of the
hydrogen atom in the parabolic coordinates \cite{12}.
Thus, we get 
\begin{eqnarray}
\psi_{n_1n_2 ms}(\xi, \eta, \varphi) = \frac{\sqrt 2}{n^2a^{3/2}}
\Phi_{n_1 m-s}(\xi)\Phi_{n_2 m+s}(\eta)
\frac{e^{im\varphi}}{\sqrt{2\pi}}, \label{6}\end{eqnarray}
where
\beq
\Phi_{p q}(x) = \frac{1}{|q|!}
\sqrt{\frac{(p+|q|)!}{p!}}
e^{-x/2an}\left(\frac{x}{an}\right)^{|q|/2}
{_1}F_1\left(-p;|q|+1; \frac{x}{an}\right).
\eeq
Here $n_1$ and $n_2$ are non-negative integers
\beq
n_1 = - \frac{|m-s|+1}{2} + \frac{ \mu}{2\kappa \hbar}\beta +
\frac{1}{2a\kappa}, \qquad n_2 = - \frac{|m+s|+1}{2} - \frac{
\mu}{2\kappa \hbar}\beta + \frac{1}{2a\kappa},
\eeq
where $\kappa=\sqrt{-2\mu E}/\hbar$. From the last relations, 
taking account (\ref{senergy}),
we get    that the parabolic
quantum numbers $n_1$ and $n_2$ are connected with the principal
quantum number $n$ as follows            
\beq
n = n_1 + n_2 + \frac{|m-s|+|m+s|}{2} + 1. \label{nn1n2}
\eeq
Thus we have solved the spectral problem 
\beq {\cal H}_{0}\psi={E}^{(0)}\psi,\quad {I}_3\psi= 
\hbar^2\beta\psi \quad
J_3\psi=\hbar m\psi, \label{sp1} \eeq 
where 
${\cal H}_{0},I_3,J_3$ 
are defined  by the expressions (\ref{1}),(\ref{2}).

\setcounter{equation}{0}
 \section{The Stark effect}
The Hamiltonian of the MIC-Kepler system in the external
constant uniform 
electric field is of the form
\begin{eqnarray}
{\cal H}=\frac{\hbar^2}{2\mu}(i{\bf{\nabla}}+ {s}{\bf A})^2
+\frac{\hbar^2{s}^2}{2\mu r^2}-\frac{\gamma}{r} +|e|\varepsilon z,
\label{8}
\end{eqnarray}
We have assumed  that the electric field ${\bf \varepsilon}$ is
directid along positive $x_3$-semiaxes,
 and the force acting the electron
is 
 directed
along
 negative  $x_3$-semiaxes.

 Since the Hamiltonian (\ref{8})
possesses axial symmetry, it is convenient  to consider the
Schr\"{o}dinger equation of the charge-dyon system in the external
constant uniform electric field in the parabolic coordinates.
Thus, for convenience, the parabolic wavefunctions (\ref{6})
of the MIC-Kepler system are chosen as nonperturbed ones
for the calculation of the matrix elements of transitions between
the mutually degenerated states.

We are mostly interested the matrix  elements of the transitions
$n_1n_2m \to n_1'n_2'm'$ for the fixed value of the principal
quantum number $n$. 
The perturbation operator in parabolic coordinates  reads
$$\hat V =
|e|\varepsilon z = |e|\varepsilon (\xi - \eta)/2.$$
 According to
the perturbation theory the first-order corrections 
to the energy eigenvalue $E_n$ (\ref{senergy}) are of the form
\beq
 E^{(1)}_n = \int\,\psi^*_{n_1n_2m}(\xi,\eta,\varphi)\hat V
\psi_{n_1n_2m}(\xi,\eta,\varphi)\,d\varphi. \label{pert}\eeq
Hence, taking into account  (\ref{6}), we  get
\beq
E^{(1)}_n = \frac{|e|\varepsilon}{4a^3n^4}\left( {\cal
I}_{n_1,m-s}I_{n_2,m+s}-I_{n_1,m-s}{\cal I}_{n_2,m+s}\right),
\label{formula}\eeq
where
\beq I_{pq} = \int_{0}^{\infty}\,\left[\Phi_{p q}(x)\right]^2\,dx,
\qquad {\cal I}_{pq} = \int_{0}^{\infty}\,x^2\left[\Phi_{p
q}(x)\right]^2\,dx. \label{cI} \eeq
Later on, we will make use the formulae (see, e.g. \cite{12})
\beq \int_{0}^{\infty}\,e^{-\lambda z}z^{\nu}
_{1}F_{1}\left(\alpha;\gamma;kz\right)\,dz =
\Gamma(\nu+1)
{\lambda^{-\nu-1}}_{2}F_{1}\left(\alpha,\nu+1;
\gamma;\frac{k}{\lambda}\right),
\eeq and 
\beq
 _{2}F_{1}\left(\alpha,\beta;\gamma;1\right)=
\frac{\Gamma(\gamma)\Gamma(\gamma-\alpha-\beta)}
{\Gamma(\gamma-\alpha)\Gamma(\gamma-\beta)}. \eeq
As a result we will obtain the following expressions 
\beq I_{pq} = an, \qquad {\cal I}_{pq} =
2(an)^{3}\left[3p(p+|q|+1)+\frac{|q|}{2}(|q|+3)+1\right]. \eeq
Now, using these formulae,
 we could get the first-order correction 
to the energy eigenvalue (\ref{senergy}) 
\begin{eqnarray}
E^{(1)}_n = \frac{3\hbar^2|e|\varepsilon}{2\mu \gamma}\left[
n\left(n_1-n_{2}+\frac{|m-s|-|m+s|}{2}\right)+\frac{ms}{3}\right].
\label{7}\end{eqnarray}
Similar to  the hydrogen atom, the linear term on $n$ is
proportional to the $x_3$-component of the Runge-Lenz vector.
However, there is an extra correction linear on $m$, which removes
 the degeneracy on the $x_3$-component of the angular momentum.

Thus, in the ``charge-Dirac dyon''
system there is the  linear Stark
effect, completely removing the degeneracy on azimuth quantum
number $m$.

For the
fixed $s$, according to the formula (\ref{nn1n2}), the two
extreme components of the splited
energy level correspond to the following
values of the parabolic quantum numbers: $n_{1}=n-|s|-1$,
$n_{2}=0$ and $n_{1}=0$, $n_{2}=n-|s|-1$.
The
distance between these  levels is
\beq
\Delta E_n = \frac{3\hbar^2|e|\varepsilon}{\mu
\gamma}n\left(n-|s|-1\right),
\eeq
i.e. the complete splitting of the level  is
proportional to $n^{2}$,
as on the case of the hydrogen atom.

The presence of the 
linear
Stark  effect means that in the
unperturbed state the charge-dyon bound system has a dipole
momentum with a mean value
\begin{eqnarray}
\overline{d}_z = -\frac{3\hbar^2|e|}{2\mu \gamma}\left[
n\left(n_1-n_{2}+\frac{|m-s|-|m+s|}{2}\right)+\frac{ms}{3}\right].
\end{eqnarray}
From the expression for the mean dipole momentum it is natural to
give the following definition for the 
dipole momentum's
operator
\begin{eqnarray}
{\bf d} = -\frac{3\hbar^2|e|}{2\mu \gamma}\left[ n^2 {\bf I}
+\frac{s{\bf J}}{3}\right]= |e|\left[ \frac{3\gamma}{4E^{(0)}}
{\bf I} -\frac{\hbar^2 s}{2\mu\gamma} {\bf J}\right] .
\end{eqnarray}
It is mentioned that all the formulae obtained for $s=0$ 
yields the
corresponding formulae for the hydrogen atom.

\setcounter{equation}{0}
\section{Conclusion}

We have considered the Stark effect in the MIC-Kepler problem
describing the interaction of charge 
with the  Dirac dyon. We have found
 that in spite of the deep similarity of the
charge-dyon system with the hydrogen atom, the relation of the
former system
to 
 the Stark effect is qualitatively differs from the latter one.
Namely,  in
the MIC-Kepler problem there is the {\sl linear} Stark effect
which completely removes the degeneracy on azimuth quantum number $m$.

It deserves to be mentioned, 
 that the 
Stark effect for the charge-dyon system
cannot be naively transferred on to the system of two
dyons. In the latter 
case the interaction of the external field
with the magnetic charge have to be  to be taken into consideration.
Hence, the Stark effect  in the  system of two Dirac
 dyons will be equivalent to the superposition of 
should be the superposition of the Stark and Zeeman
effects in the charge-dyon system.

Undoubtedly, the existance in the charge-dyon system 
of the linear  Stark  effect, as well as of the  nonzero dipole
momentum is due to the presence of magnetic monopole.
It is clear, the  system of two well-separated BPS dyons 
will  has similar properties, thought 
the transformation of the  Schr\"{o}dinger
equation corresponding to  the Hamiltonian (\ref{8}) yields  the
Schr\"{o}dinger equation of the system of two well-separated BPS dyons,
perturbed by a nonlinear electric field. Besides, the
formal equivalence of the  MIC-Kepler system   and of the two well-separated
 BPS dyons supposes the dependence of the coupling 
 constant of the latter
system both on  the coupling  constant and of the energy of the
MIC-Kepler.

It seems to be  interesting consider  the Stark effect 
in the MIC-Kepler system on the sphere and
hyperboloid, with the aim of the  revealing its dependence on the space
 curvature. The study of  the Stark effect on 
the five-dimensional generalization of the
MIC-Kepler problem (the so-called Yang-Coulomb or $SU(2)$-Kepler
problem) \cite{iwa1} could be even  more instructive, 
due to the presence of isospin degrees of freedom.
\vspace{5mm}

{\large Acknowledgements.} The authors are grateful to Valery
Ter-Antonyan for useful discussions and remarks.
This work was partially
supported   by  ANSEF No: PS81 and INTAS 00-00262  grants.

\end{document}